\documentclass[journal]{IEEEtran}
\usepackage[utf8]{inputenc}
\usepackage{times}
\usepackage[cmex10]{amsmath}
\usepackage{amsfonts,amssymb}
\usepackage{array}
\usepackage{graphicx}
\usepackage[usenames,dvipsnames]{xcolor}
\usepackage[linesnumbered,ruled,vlined]{algorithm2e}
\usepackage{soulutf8}
\usepackage{xcolor}
\usepackage{color}
\usepackage{cite}
\usepackage{booktabs}
\usepackage{subfigure}
\usepackage{siunitx}
\usepackage{multicol}
\usepackage{array}
\usepackage{url}
\newcolumntype{P}[1]{>{\centering\arraybackslash}p{#1}}

\begin{document}

\title{Improving 6TiSCH Reliability and Latency with Simultaneous Multi-Band Operation}

\author{Marcus Vinicius Bunn, Richard Demo Souza, and Guilherme Luiz Moritz 
        \thanks{M. V. Bunn and R. D. Souza are with the Federal University of Santa Catarina, Brazil, marcus.bunn@posgrad.ufsc.br, richard.demo@ufsc.br.}  \thanks{G. L. Moritz is with the Federal University of Technology-Paraná, Brazil. moritz@utfpr.edu.br.} 
        \thanks{This work has been partially supported in Brazil by CAPES, Finance Code 001, PrInt CAPES-UFSC ``Automation 4.0'' and CNPq.}}

\maketitle

\begin{abstract}The Internet Engineering Task Force (IETF) group "IPv6 over the TSCH mode of IEEE 802.15.4e" (6TiSCH) introduced a protocol, utilizing Time-Slotted Channel Hopping (TSCH) from IEEE802.15.4e due to its high reliability and time-deterministic characteristic, that achieves industrial performance requirements while offering the benefits of IP connectivity. This work proposes the addition of a second radio interface in 6TiSCH devices to operate a parallel network in sub-GHz, introducing transmit diversity while benefiting from decreased path-loss and reduced interference. Simulation results show an improvement of 25\% in Packet Delivery Ratio (PDR) and closely to 30\% in latency in different 6TiSCH networks scenarios.  
\end{abstract}


\section{Introduction}
\label{sec:introduction}

{T}{he} Industry 4.0 paradigm promises unprecedented improvements in productivity, control, maintenance and cost reduction to factories and industries, while enabling the development of new products and processes~\cite{drath2014}. One of the main technologies supporting this evolution is the Industrial Internet of Things (IIoT)~\cite{hermann2016}, enhancing factory connectivity levels, powering varied applications and integrating Wireless Sensor Networks (WSN) with the Internet. A major challenge is to guarantee the communication requirements in terms of determinism, latency and reliability for critical industrial applications~\cite{vitturi2013}. For many years the increased reliability of wired networks has suppressed the benefits of mobility, flexibility and cost reduction of wireless networks~\cite{gungor2009}, leaving wireless deployments for secondary systems~\cite{candell2017}. 



A set of industrial communication protocols have been designed to address the above challenges, as WirelessHART~\cite{chen2010}, ISA100.11a~\cite{isa1002009} and WIA-PA~\cite{wiapa2015}. These protocols are based on Time-Slotted Channel Hopping (TSCH) mode of IEEE802.15.4e~\cite{ieee2015}, due to its high reliability and time-determinism. By delivering 99.999\% end-to-end reliability and over a decade of battery lifetime~\cite{watteyne2015}, TSCH has become the \textit{de-facto} Medium Access Control (MAC) technique for industrial applications~\cite{yu2014}. Moreover, the continuous increase in prediction of future connected devices, and the proven applicability of the IIoT to fulfill the Industry 4.0 requirements~\cite{ieeei40}, encouraged the creation of the Internet Engineering Task Force (IETF) group ``IPv6 over the TSCH mode of IEEE 802.15.4e'' (6TiSCH). Their efforts resulted in a communication protocol capable of achieving industrial performance requirements while offering the benefits of IPv6 connectivity~\cite{vilajosana2019}. 

However, wireless communication in rough industrial environments remains challenging, due to interference and fading. Interference may be caused by other technologies, by a secondary deployment, or between devices in the same network, while fading is inherent to the wireless link~\cite{dujovne2014}. TSCH increases network performance over multi-path fading and interference, but the continuous increase in connected devices combined with the strict reliability and latency requirements of the Industry 4.0 paradigm~\cite{lu2016} pose new challenges. Therefore, existing and continuous efforts from the industry and academia are required to improve IIoT networks performance.

Some related work aim at improving TSCH via redundant transmissions~\cite{minet2017,papadopoulos2017} and the usage of sub-GHz band with multi-band support~\cite{yin2014,brachmann2019}. However, it is noticeable the absence of a single approach that combines both methods, and which can improve TSCH network performance against interference and multi-path fading. In this context, this work proposes the addition of a second radio interface in 6TiSCH devices to operate a parallel network in sub-GHz band, which increases the network reliability, latency, and connectivity by introducing transmission diversity while benefiting from decreased path-loss propagation and reduced interference from other technologies. The results of several simulations show potential improvements of up to 25\% in Packet Delivery Ratio (PDR) and closely to 30\% in latency in different tests, at the cost of increased hardware complexity.

This paper is organized as follows. Section~\ref{sec_6tisch} presents a brief 6TiSCH overview, followed by a discussion of related works in Section~\ref{sec_related}. The proposed configuration is exposed in Section~\ref{sec_config} and Section~\ref{sec_methodology} details the methodology applied in the performance evaluation. Section~\ref{sec_results} discusses numerical results, while Section~\ref{sec_conclusions} concludes the paper.

\section{6TiSCH Overview}
\label{sec_6tisch}

The IETF IPv6 over the TSCH mode of IEEE802.15.4e (6TiSCH) Working Group (WG) has been established to produce specifications of an interoperable stack integrating IEEE802.15.4e TSCH to IETF solutions targeting IoT applications~\cite{vilajosana2019}.  The stack uses the IEEE802.15.4 physical layer (PHY) operating in the 2.4 GHz (ISM) band. This band is divided in 16 channels~\cite{ieee2015} whose use is governed by the Time Slotted Channel Hopping (TSCH) IEEE802.15.4e mode, which combines Time Division Multiple Access (TDMA) with channel hopping to create a collision free environment which can increase reliability over multi path fading and interference. Adittionally, 6TiSCH provides a set of management protocols that enables plug-and-play bootstrap, authentication and wireless medium management~\cite{I-D.ietf-6tisch-minimal-security,I-D.ietf-6tisch-msf,RFC8137,RFC8180,RFC8480}.

\begin{figure}[!t]
\center{
 \includegraphics[width=\columnwidth]{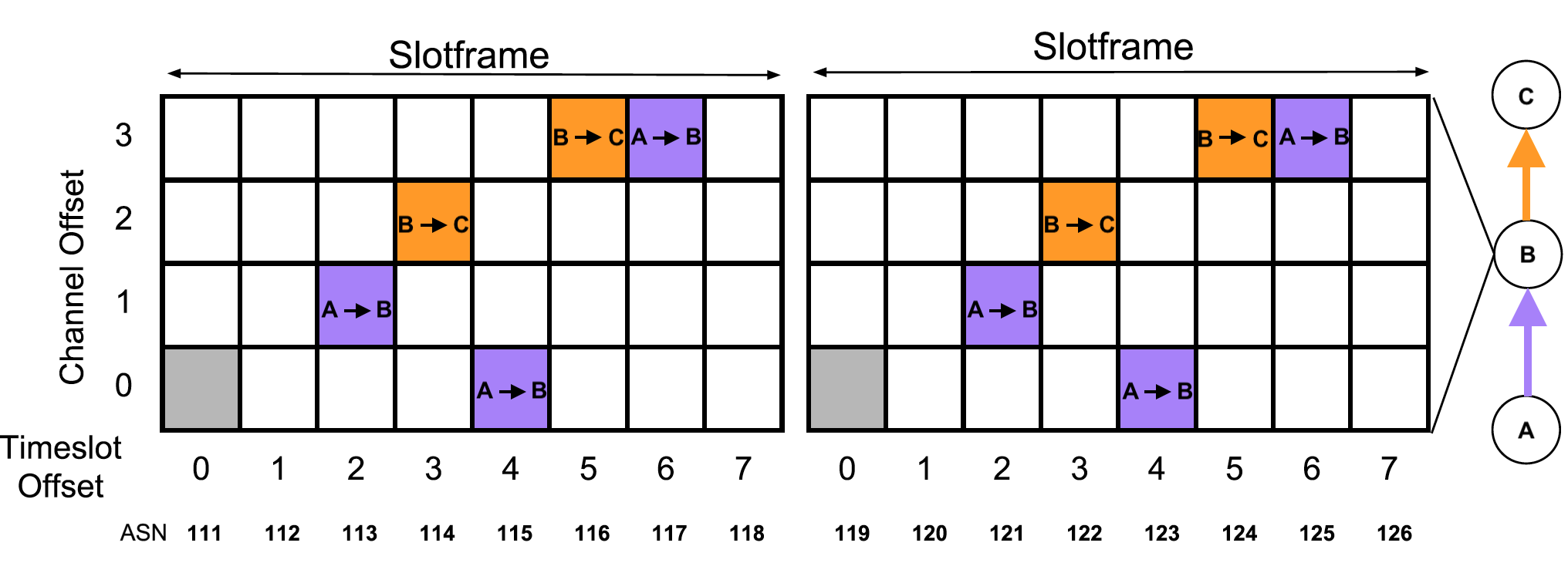}}
\caption{The 6TiSCH schedule example. The depicted slotframes contain 7 timeslots with 4 available physical channels and two cycles are shown. Cells are mapped to a execution time and channel frequency based on time (0 to 7) and channel (0 to 3) offsets. Gray timeslots are reserved for broadcast while Purple and Orange timeslots represent unicast communication. Purple timeslots are dedicated links for Node A to Node B transmissions. Orange ones are dedicated for Node B to C transmissions. Adapted from~\cite{vilajosana2020}.\label{slotframe}}
\end{figure}

In the 6TiSCH stack, communication occurs in specific times while obeying a maximum duration determined by a \textit{timeslot}. Timeslots repeat in time indefinitely and a group of timeslots is named a \textit{slotframe}. A scheduling function determines whether a node is transmitting, receiving or sleeping in each timeslot, which can offer deterministic and reliable communication with improved battery lifetime by allowing nodes that are not transmitting or receiving to enter in sleep mode. The resulting allocation, named schedule, can be viewed as a repeating $M\times{}N$ matrix, where $M$ is the number of available physical channels and $N$ is the slotframe length, as depicted in Figure~\ref{slotframe}. Channel hopping is achieved by selecting offsetting channel cells in each slotframe iteration~\cite{vilajosana2020}.

To provide a zero configurarion network, the 6TiSCH minimal configuration~\cite{RFC8180} defines a mandatory basic schedule which must be followed by any 6TiSCH node. This minimal schedule provides basic message exchange that can be used in conjunction with the 6TiSCH Operation Sublayer (6top) Protocol (6P)\cite{RFC8480} to negotiate more complex communication schedules governed by a Scheduling Function (SF). {A mandatory basic Schedule function, named Minimal Scheduling Function (MSF)~\cite{I-D.ietf-6tisch-msf} is provided by 6TiSCH.} 



After single link communication is established, routing is provided by RPL~\cite{RFC6550}, which defines four types of messages to create a directed acyclic graph (DAG) towards the root. A new node that wants to join the network (named pledge) must listen to messages carrying DODAG joining information (DIO) which are periodically broadcast by the network and can also be actively solicited by the pledge using DODAG information solicitation messages (DIS). Upon receiving DIO messages from multiple nodes, a pledge uses an Objective Function to choose a parent that will be the first hop of all the pledge messages. This function is implemented to achieve specific requirements, as to increase network lifetime and avoid loops~\cite{pereira2020}. To complete the joining procedure, the selected parent must be addressed with a destination advertisement object (DAO) message which is replied by a DAO acknowledge (DAO-ACK) in the case of a joining accept.

At the transport and network layers, 6TiSCH stack uses IPv6 over Low-Power Wireless Personal Area Networks (6LoWPANs)~\cite{RFC4919} to compress User Datagram Protocol (UDP) and IPv6 headers. In the application layer it uses Constrained Application Protocol (CoAP)\cite{RFC7252}, secured by a tool called Object Security for Constrained RESTful Environments (OSCORE)\cite{RFC8613}. The 6TiSCH minimal configuration defines the Constrained Join Protocol (CoJP) for a secure joining process. The process is executed in a single transaction, where the pledge sends authentication data to one available neighbor, named Join Proxy (JP) that is forwarded to the Join Registrar/Coordinator (JRC). Transactions are executed after RPL join process is initiated and encrypted keys are sent over the minimal cell for authentication. If approved, the JRC notifies the pledge to confirm its addition to the network.


\section{Related Work}
\label{sec_related}
This section discusses related work that use the TSCH mode of IEEE 802.15.4e and that propose redundant transmission, sub-GHz operation and multi-band support. 

Minet et al.~\cite{minet2017} exploit redundant transmissions that benefit from different communication links to increase reliability, where a node sends a message through multiple paths depending on a redundancy pattern. The sink node accepts the first delivered message and discards the late copies, which increases reliability and reduces latency. The increase in reliability is achieved at the cost of additional network overhead that decreases battery lifetime. Moreover, additional studies would be required to confirm the effectiveness of the proposed strategy on interference prone environments with coexisting networks, where redundant transmissions could degrade performance by increasing network density and interference levels. 

Papadopoulos et al.~\cite{papadopoulos2017} 
propose redundant transmissions associated with an overhearing mechanism to increase reliability and reduce latency. Each node forwards its messages not only to the default RPL parent but also to a redundant parent. In addition, packet retransmissions due to incorrect receptions are eliminated. Simulation results were compared against the default TSCH-RPL network using different retransmission levels, demonstrating a reduction of up to 54\% in end-to-end latency and 84\% in jitter when compared with a non redundant scenario with 8 retransmissions at the cost of increased energy consumption caused by the redundant transmissions. Regarding PDR, results showed no improvements when compared to the the retransmission approaches, and the authors justify this behavior by stating that the removal of retransmissions negatively impacted the control packets reliability. 

Yin, et al.~\cite{yin2014} tackles the interference problem on WSNs that operate in the 2.4 GHz band caused by popular WiFi and Bluetooth network deployments by proposing dual band operation. The scheme performs sequential transmissions for both 900 MHz and 2.4 GHz. Experiments were conducted to evaluate the proposed scheme performance on two different testbeds~\cite{twonet2013,opal2011}. The PDR is selected as evaluation metric and tests are executed over varied wireless channels from 900 Mhz to 2.4 GHz. Results show that the average PDR was approximately 5\% higher in the 900Mhz band, while also improving the connectivity by 15\%, when compared to the 2.4 GHz band. It concludes, based on experimental results, that the presented scheme can be used to increase network performance and connectivity, although the paper focuses only on the physical/link layers. 


Brachmann, et al.~\cite{brachmann2019} propose multiple frequency and bitrates in a single IEEE 802.15.4e TSCH schedule to meet multiple application requirements by trading datarate with robustness. Two approaches are investigated, the first assigning  timeslots duration to accommodate the slowest transmission and the second  allowing slower transmissions to use several timeslots. The performance of the proposed schemes were evaluated experimentally using 25 nodes deployed in an office environment. For the tests, TSCH control data was transmitted in the sub-GHz band that offer increased reliability while application data is transmitted over 2.4 GHz to achieve faster delivery times. 
The usage of sub-GHz bands granted single-hop reaches close to 24 nodes at 1.2kbps, while at the standard 250kbps in 2.4 GHz the reach decreases drastically to an average of 10 nodes. 
Results also showed that the 1.2 kbps band at sub-Ghz has a 20x higher channel utilization when compared to the 2.4GHz band at 1000 kbps, while improving network synchronization by reducing the required average hops for control data. The work successfully demonstrates the required timing configuration required in TSCH networks to operate in sub-GHz and allows multi-band operation. 




Against the above background, the solution proposed in this paper can be seen as an extension of~\cite{yin2014} and~\cite{brachmann2019} by exploiting coexisting 2.4 GHz and sub-GHz networks while combining them at the application level to improve the overall network performance. Additionally, it differentiates from~\cite{minet2017} and~\cite{papadopoulos2017} by applying redundant transmissions and exploiting diversity in a more ``standard'' fashion, where redundancy occurs naturally by using the additional operating band combined with frequency and spatial diversity associated with the different TSCH and RPL networks.

\section{Proposed Configuration}
\label{sec_config}
This work proposes multi-band support, including 2.4GHz and sub-GHz bands, in 6TiSCH networks by employing two independent radio interfaces with concurrent transmissions. This section describes how this scheme can be implemented and configured in 6TiSCH networks. 

\subsection{Sub-GHz Operation}

Industrial environments may include several applications, each having different requirements that can be met by a variety of wireless technologies. The dependability of wireless communications can be associated with specific criteria, for example regarding availability and reliability, where communication must be executed when needed while respecting a time-limit requirement~\cite{fukalas2019}. In other words, reliability is not only related with the correct reception of a packet but with its validation, where a late packet can become obsolete, {\it e.g.} a fire alarm event. However, other applications are more tolerant to packet delivery delay, and can benefit from more robust communication link and longer battery lifetime. 

Standalone usage of sub-GHz band for industrial applications is commonly associated with monitoring and remote communication due to its end-to-end time delivery limitation and strong signal propagation~\cite{candell2018}. For instance, LPWANs rely on sub-GHz bands, simple MAC protocols and star network topologies to offer massive connectivity and wide coverage~\cite{raza2017}, e.g LoRaWAN~\cite{sornin2017}. On the other hand, industrial focused technologies, such as IEEE 802.15.4e TSCH, are specified towards reliable and deterministic communications powered by complex MAC and scheduling protocols. 


Introducing sub-GHz bands along the default 2.4 GHz frequencies offers longer communication range and less interference with co-existing technologies~\cite{brachmann2019}. By allowing multi-band support, the resulting deployment can, for example, improve network performance due to frequency diversity and serve different applications with a variety of requirements. Our focus sits on the first, to improve performance by introducing diversity to communication links in different operating bands. 


\subsection{Multi-band Support}
This section presents the required changes to support the additional operating band in a 6TiSCH network while making them coexist as a single network at the application layer. First, as mentioned, 6TiSCH is designed to follow the IEEE 802.15.4e Physical Layer (PHY) at $2.4$GHz with 16 channels, each spaced by 5MHz, and transmit at 250kbps rate. At this rate, within a 10ms TSCH timeslot it is possible to transmit a data frame and to receive an acknowledgment~\cite{brachmann2019}.


To allow operation in sub-GHz bands and multi-band support we follow the IEEE 802.15.4g-2012 standard~\cite{ieee2012}. More precisely, we select  Operating Mode \#1 for the 863-870MHz band in Europe while Table~134 in~\cite{ieee2012} lists the following parameters: 50kbps transmission rate and 200kHz channel spacing in a total of 34 available channels. {In this band, given the absence of a TSCH standardized value for timeslot duration, we chose $29.38$ms as in~\cite[Table III]{brachmann2019}, mainly because of their proven efficiency and thorough tests.}


In this context, the three main required changes in 6TiSCH to enable sub-GHz bands and to support multi-band simultaneous operation are: 1) increase timeslot duration in the sub-Ghz band; 2) redefine TSCH channels in sub-GHz (34 channels instead of 16 in 2.4GHz); and 3) application data combining at the network sink (DAG root). {With this configuration, a packet is scheduled for transmission as soon as it is generated. The actual transmit time may be different for each interface due to the independence between timeslot duration and scheduling configurations in different bands.}  A proposed application and its required configuration to account for multi-band operation will be presented next.





\subsection{Performance Metrics}

In the flowing sections, some performance metrics that are used to ensure reliable, deterministic and time-sensitive communication for industrial applications are defined.

\subsubsection{PDR} 

\begin{figure}[!t]
\center{
 \includegraphics[width=0.8\columnwidth]{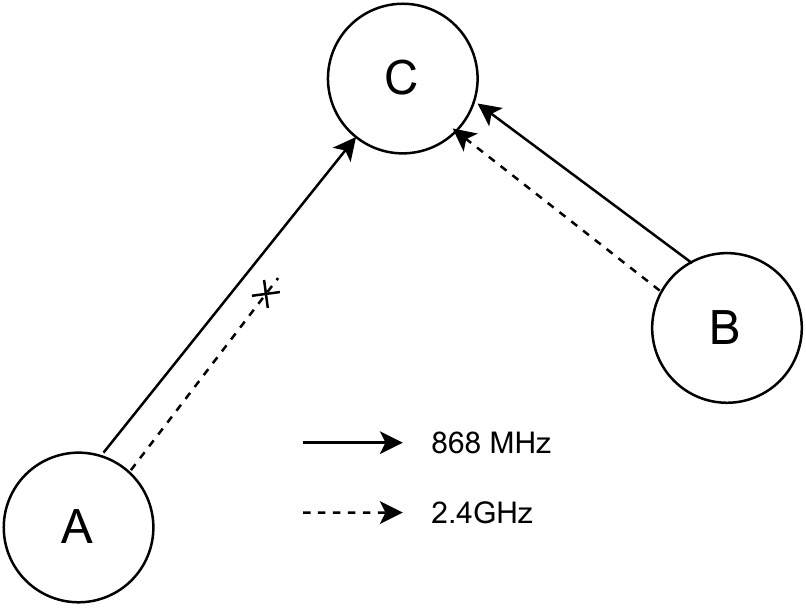}}
 \caption{Node A distance to its parent C inflicts greater path-loss and higher packet loss probability when compared to Node B. Packets are transmitted simultaneously in both radio interfaces where only the packet from the sub-GHz band is effectively decoded at C. The network benefits from the transmission diversity introduced by the sub-GHz interface. \label{fig_multi_1}}
\end{figure}

Consider $\mathcal{S}$ a set containing $|\mathcal{S}|=N$ nodes in a simulation run. Next, consider $\mathcal{M}^{\text{Tx}}$ the set of generated messages by all nodes $\in \mathcal{S}$. Each packet $p \in \mathcal{M}^{\text{Tx}}$ is transmitted to the root node using all the available interfaces of the generating node. To be considered correctly received, $p$ must arrive correctly in at least one of the DAG root wireless interfaces. A packet may be lost due to effects of the wireless medium, consequently, only a subset of $\mathcal{M}^{\text{Tx}}$, denoted $\mathcal{M}^{\text{Rx}}$ is correctly received. The network PDR can be defined as: 
\begin{equation}
\text {PDR} = \frac{|\mathcal{M}^{\text{Rx}}|}{|\mathcal{M}^{\text{Tx}}|}.
\end{equation}
\noindent
Transmitting using multiple interfaces at different frequencies increases the PDR since each packet follows a different path, having a  lower probability of being in outage simultaneously than each one individually, as illustrated in Figure~\ref{fig_multi_1}.

\noindent


\subsubsection{Latency}

Consider $t_p^{\text{Tx}}$ the time when a packet $p \in \mathcal{M}^{\text{Rx}}$ is generated at node $i \in \mathcal{S}$, while $t_p^{\text{Rx}}$ is the time when this same packet $p$ firstly arrives at any of the DAG root interfaces. The average network latency $L$ is
\begin{equation}
L = \frac{\sum_{ p \in \mathcal{M}^{\text{Rx}}} \left[ t_p^{\text{Rx}} - t_p^{\text{Tx}} \right] }{|\mathcal{M}^{\text{Rx}} |}.
\label{eq_latency_single}
\end{equation}

\begin{figure}[!t]
\center{
 \includegraphics[width=0.8\columnwidth]{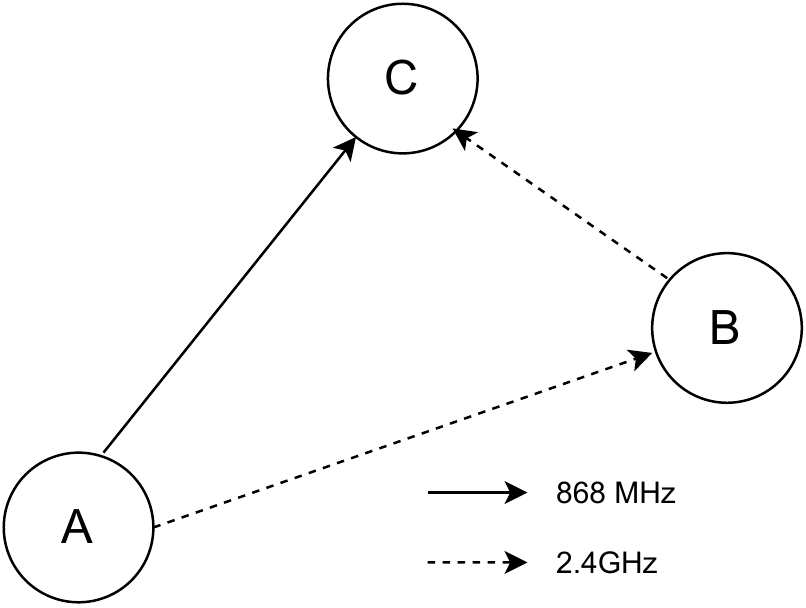}}
 \caption{Node A simultaneously sends a  packet by both interfaces, with different RPL topologies and TSCH scheduling. In the $2.4$Ghz interface, Node A transmits the packet to its parent B, which then relays to the destination C. In the $868$Mhz interface, the packet is directly transmitted from A to  C. Sub-GHz bands allow longer hops, reducing latency by eliminating multiple scheduling and transmission processes in packet forwarding. \label{fig_multi_2}}
\end{figure}

Figure~\ref{fig_multi_2} illustrates the benefit in latency that can be introduced by the usage of sub-GHz bands when the amount of average hops is decreased. Additionally, revisiting Figure~\ref{fig_multi_1} it can be noticed how latency can also be improved by reducing packet retransmissions, where a successful communication is executed by the sub-GHz band while the $2.4$GHz would require multiple retries to complete the transmission. 

\subsection{Contribution}

We propose multi-band support with sub-GHz operating bands in conjunction to the default 2.4 GHz band in order to increase reliability and reduce latency. This effect can be achieved by introducing frequency diversity, reduced interference and increased robustness. By using diverse paths to the DAG root, we envision that it is possible to improve the overall performance in terms of PDR and latency, simultaneously.

\section{Methodology} \label{sec_methodology}

This section describes the methodology employed in the experiments conducted in this study. Next, the simulation environment, the 6TiSCH Simulator configuration, the simulation scenario, and the applied performance metrics are presented.

\subsection{Simulation Environment}

The 6TiSCH Simulator from~\cite{municio2019} was used for obtaining experimental results. This simulator was created by researchers from the working group that developed 6TiSCH. In addition to the default TSCH configuration for the $2.4$GHz operating band and propagation model based on OpenMote~\cite{vilajosana2015openmote} already available, a second sub-GHz configuration was added, based on the Texas Instruments CC1352R~\cite{ti2018} radio operating in the 802.15.4g SUN PHY at $868$MHz. Table~\ref{tab_scenario} summarizes the configuration of the simulation environment.

\subsection{6TiSCH Configuration}
Few changes from the default 6TiSCH Simulator version 1.3.0~\cite{simulator2020} configuration were made to allow the usage of the 868 Mhz band and are described bellow. First, packet generation interval, $T_a$, was changed to 10 seconds in order to increase the amount of packets traveling in the network, and as a consequence, to increase collision and packet loss probability. Second, uniform variance, $V$, applied to the packet creation time, was set to zero to force equal packet creation times for both radio interfaces. Third, the retransmission number, $R_t$, was configured as the recommended maximum number of link-layer retransmissions defined in RFC 8180~\cite{RFC8180}. Fourth, the clock drift, $C_d$, between a device and its time reference neighbor was set to zero to establish a controlled environment to evaluate the proposed configuration. Lastly, to account for the different network formation times in both operating bands, application packets transmissions were scheduled to wait a fixed time and start transmitting simultaneously in both interfaces after the network creation process time is completed.

\subsubsection{Propagation Model}

The 6TiSCH Simulator was designed by the 6TiSCH IETF working group with three major goals: compliance with the standard, scalability and simplicity. Its propagation model is based on the Pister-Hack~\cite{pister2009} model, where the path loss is uniformly distributed between that obtained from the Friis equation (free space) and Friis plus 40 dB. The resulting values are used as a link quality metric in further operations, such as DAG formation and for calculating packet reception probability~\cite{municio2019}. To determine reception probability, RSSI values are first converted to PDR values based on a conversion table. The table was obtained empirically in a real deployment utilizing the OpenMote devices. According to~\cite{brun2016} the conversion table accurately reflects the relationship between  RSSI and PDR in large indoor industrial scenarios at the $2.4$GHz band. 

Therefore, to account for a different operating band a new conversion table is required, which is constructed by applying an offset to the original one based on the differences in radio sensitivity from each operating band devices. More specifically, it was applied a 13 dB offset regarding the difference from the default Texas Instruments CC2538 radio~\cite{ti2012} sensitivity to that of the Texas Instruments CC1352R.

Morevoer, the interference modeling considers the signal-to-interference-plus-noise ratio (SINR), which is calculated by adding the RSSI from the interfering neighbors to the background noise. Then, the SINR is converted to PDR values and compared to a random number to determine if the reception was successful.

\subsection{Simulation Scenario}

To evaluate the multi-band support and sub-GHz usage, a simulation scenario with 2 different bands of operation and 12 topologies were implemented. Network topologies are formed by 3 different network sizes $N \in \{40, 80, 160\}$, and 2 deployment models, namely Linear and Random. 

Nodes collect data periodically and transmit them to the DAG root abiding by each band TSCH scheduling and RPL configuration. MSF determines when communication occurs for each node to its neighbor at every hop in the RPL DAG. If no cells are available for one node to communicate with its neighbor, the transmission is scheduled for the next slotframe, where MSF will control if additional communication timeslots are required. The simulation runs twice, first for 2.4GHz band and then for 868 MHz band. Finally, metrics are combined to analyze the resulting network peformance.

The simulator uses a 2D square surface of side $L$ to define the deployment area, where Cartesian coordinates are used to define device positioning. In the Linear topology, nodes are at equal distance from their neighbors in both directions, except for the DODAG root which is positioned at the center in coordinates ($L/2$, $L/2$). In the Random topology, nodes are positioned randomly and the only requirements are: 1) the DODAG root is at the center; 2) all nodes must be within the area; and 3) nodes must have at least one reachable neighbor.


\begin{table}
\caption{Simulation scenario parameters.}
\centering
\setlength{\tabcolsep}{3pt}
\begin{tabular}{|l|c|}
\hline
Parameters & Values \\
\hline
Network size & 40-80-160 nodes\\
Area & 100mx100m \\
Setup time ($W$) & 5400s \\
Duration time & 7200s \\
Random seed & 0x74C2A74018BDB\\ 
Message interval ($T_a$) & 10s (uniform) \\
UDP payload & 90 bytes \\
Retranssmissions ($R$) & 3\\
\hline
\end{tabular}
\label{tab_scenario}
\end{table}



\section{Experimental Results} \label{sec_results}

This section presents the results of 24 simulated experiments\footnote{The source code used in this work is available at~\cite{bunn2020}.}. The discussion regarding PDR and latency, for each individual interface and for the resulting multi-band network, is presented in the following subsections.




\subsection{Packet Delivery Ratio}

\begin{figure}[!t]
\center{
 \includegraphics[width=\columnwidth]{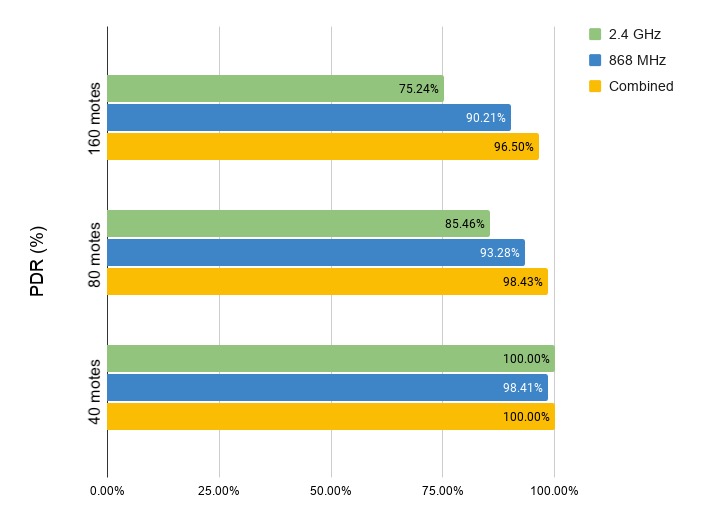}}
 \caption{PDR results for each operating band varying in network size with 40, 80 and 160 nodes randomly deployed in a 10km\textsuperscript{2} area. \label{fig_pdr_ran_deploy}}
\end{figure}

\begin{figure}[!t]
\center{
 \includegraphics[width=\columnwidth]{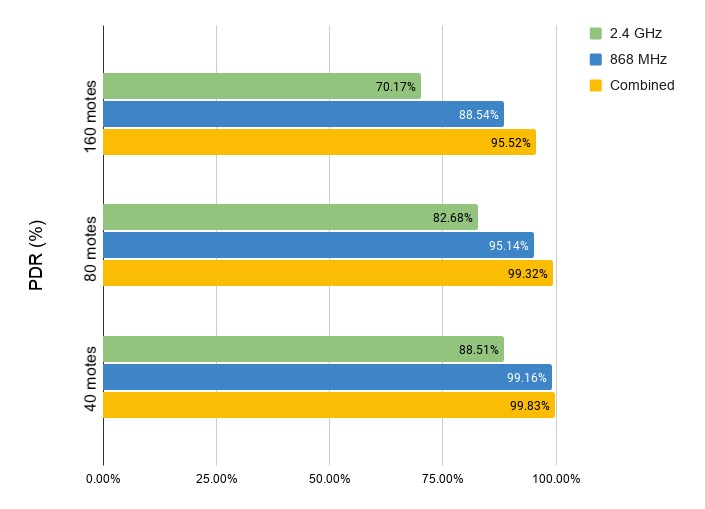}}
 \caption{PDR results for each operating band varying in network size with 40, 80 and 160 nodes linearly deployed in a 10km\textsuperscript{2} area.\label{fig_pdr_lin_deploy}}
\end{figure}

We initiate our discussions by first presenting results concerning the PDR for each combination of network size, operating band and deployment in Figures~\ref{fig_pdr_ran_deploy} and~\ref{fig_pdr_lin_deploy}. In addition to that, the associated joint metric resulted from the combination of both interfaces is also presented. It can be noticed that the PDR decreases with the network size, effect that is most significantly observed in the $2.4$Ghz band. The network size increase also degrades the joint metric results from multi-band support, yet the proposed configuration still considerably improves the overall performance. The most significant improvement is observed in the Linear deployment with 160 devices as demonstrated in Figure~\ref{fig_pdr_lin_deploy}. There, multi-band support improved PDR by 25\% when compared to the single usage of $2.4$GHz and close to 15\%  when compared to the single usage of $868$MHz band. Similarly, multi-band support offered an increase of 11.32\% and 16.64\% in the PDR for the 40 and 80 network sizes using Linear deployments. 

Regarding Random deployments, in most cases the same behavior was observed. The increase in network size resulted in lower network performance, while the multi-band support yielded significant improvements. However, Figure~\ref{fig_pdr_ran_deploy} shows that the 40 device network behaved differently, where the 2.4 GHz band achieved better PDR when compared to the 868 MHz band. Further investigation showed that the major cause for packet losses are the packet drops in the device TSCH queue. The reason for that is twofold: 1) network density; and 2) unbalanced network load. The former is associated with an increase of packet transmissions, and as a consequence, the increase in network density causing congestion in the TSCH queue. The latter can be observed in nodes near the DODAG root, which have a higher probability of being the preferred parent by multiple nodes. This unbalanced network overload can create bottlenecks that result in packet drops~\cite{farag2020}. 
The longer communication links in the 868 Mhz band, and therefore fewer transmission hops, increase the probability of unbalanced RPL topologies with congested queues. 

Moreover, the reason for decreased performance over larger networks is that the more denser the network, the higher is the interference and the strain over  bottlenecks nodes closer to the DAG root~\cite{farag2020}. This loss in performance is most noticeable in $2.4$GHz operating bands mainly due to its weaker sensitivity.

\subsection{Latency}

\begin{figure}[!t]
\center{
 \includegraphics[width=\columnwidth]{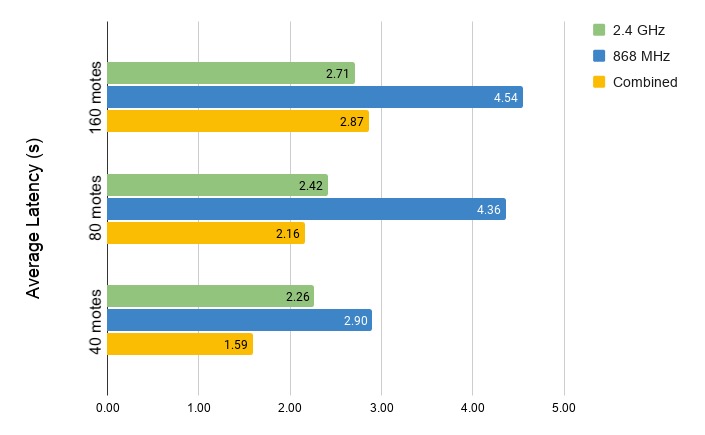}}
 \caption{Average latency results for each operating band varying in network size with 40, 80 and 160 nodes linearly deployed in a 10km\textsuperscript{2} area. \label{fig_lat_lin_deploy}}
\end{figure}

\begin{figure}[!t]
\center{
 \includegraphics[width=\columnwidth]{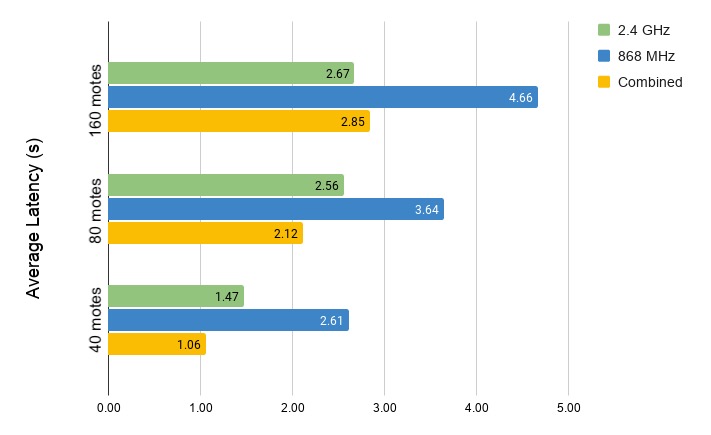}}
 \caption{Average latency results for each operating band varying in network size with 40, 80 and 160 nodes randomly deployed in a 10km\textsuperscript{2} area. \label{fig_lat_ran_deploy}}
\end{figure}

In terms of latency, similarly to the case of PDR, the increase in network size resulted in poorer overall network performance, while the multi-band support yielded significant improvements. Figures~\ref{fig_lat_lin_deploy} and~\ref{fig_lat_ran_deploy} present the average latency for each operating band and the resulting joint  metric in case of multi-band support. It can be noticed an improvement of 29.65\% and 6.61\% in the average latency by combining both operating bands when deploying Linear networks of size 40 and 80, respectively. Similar behavior was observed in the Random deployment, obtaining an improvement of 27.89\% and 17.25\% in the average latency with 40 and 80 nodes. This benefit is associated with reduced packet retransmissions and reduced average hop number in packet forwarding.

However, as shown in Figure~\ref{fig_lat_lin_deploy}, one can observe a 6.32\% degradation in the average latency for the joint metric when compared to the 2.4 GHz band in the 160 nodes scenario. This result is related to the increase in PDR provided by multiband operation. In the combined network, the additional received packets were transmitted via the 868 Mhz band that has an increased latency due to its longer timeslot to acco. While the increase in average latency can appear to be harmful to the network, the increase in successfully received packets offered by combining both bands is essential to the correct execution of certain applications, thus representing an appealing trade-off. A similar behavior can be observed in the Random deployment as illustrated in Figure~\ref{fig_lat_ran_deploy}.

\subsection{Discussion}

The experiments showed that multi-band support is beneficial for 6TiSCH networks and various industrial applications by providing frequency diversity and reducing interference from other technologies, thus increasing PDR. Also, multi-band support is useful in decreasing average packet retransmissions, hence allowing lower end-to-end latency in most cases. Regarding latency, it can be noted that duplicated packets have different propagation channels and face distinct routing paths as a consequence of the different DAGs for each band. {Single hop transmissions in the 2.4 GHz band offer the fastest transmission time, as a consequence to its higher transmission rate and shorter slotframe duration. On the other hand, due to its lower robustness and shorter range, 2.4GHz interfaces have a greater chance of requiring retransmissions or a higher number of hops to reach the destination.}

\begin{figure}[!t]
\center{
 \includegraphics[width=\columnwidth]{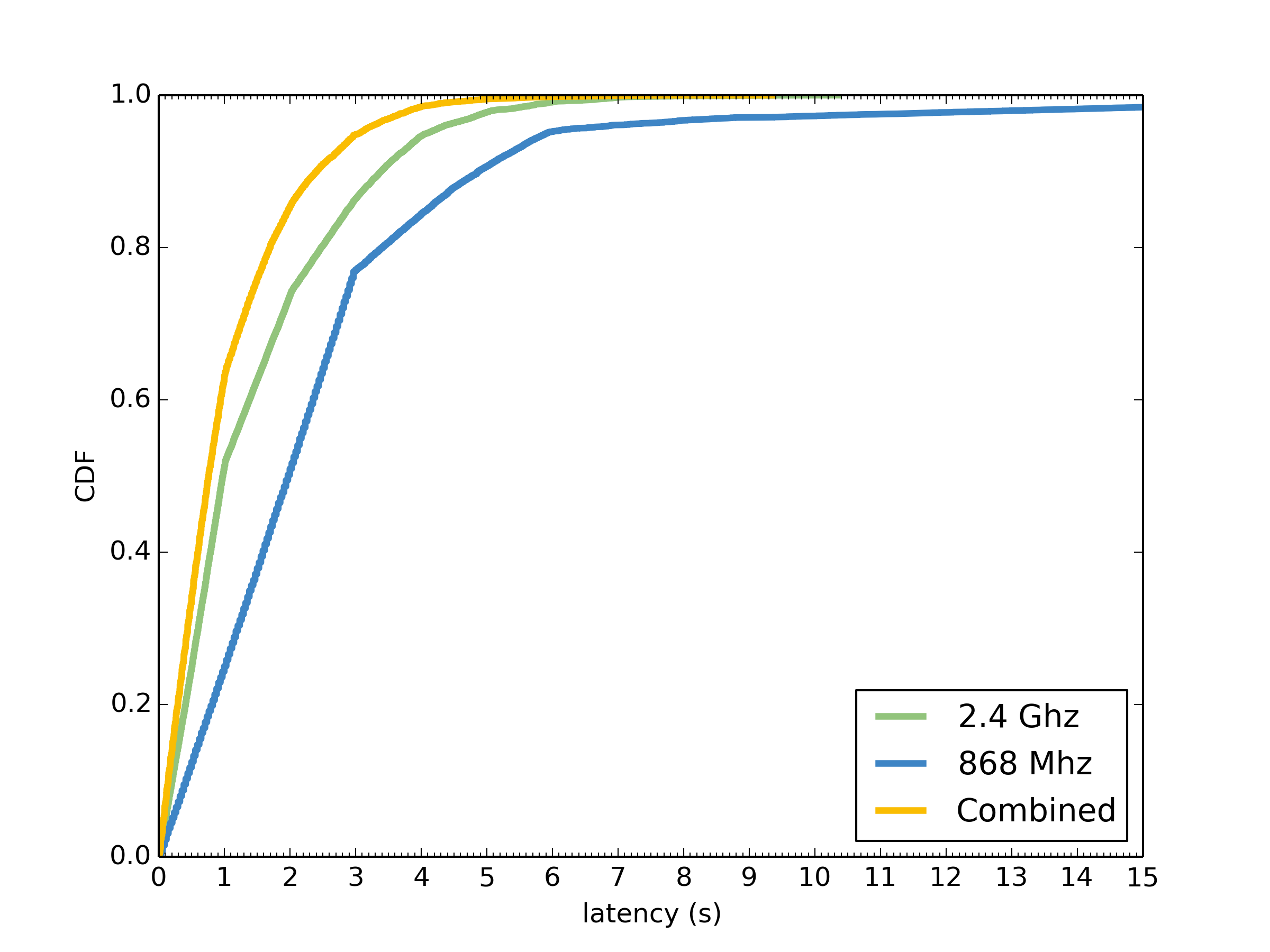}}
 \caption{Latencies CDF for the scenario with 40 nodes randomly deployed. \label{fig_cdf_lat_40_ran}}
\end{figure}

\begin{figure}[!t]
\center{
 \includegraphics[width=\columnwidth]{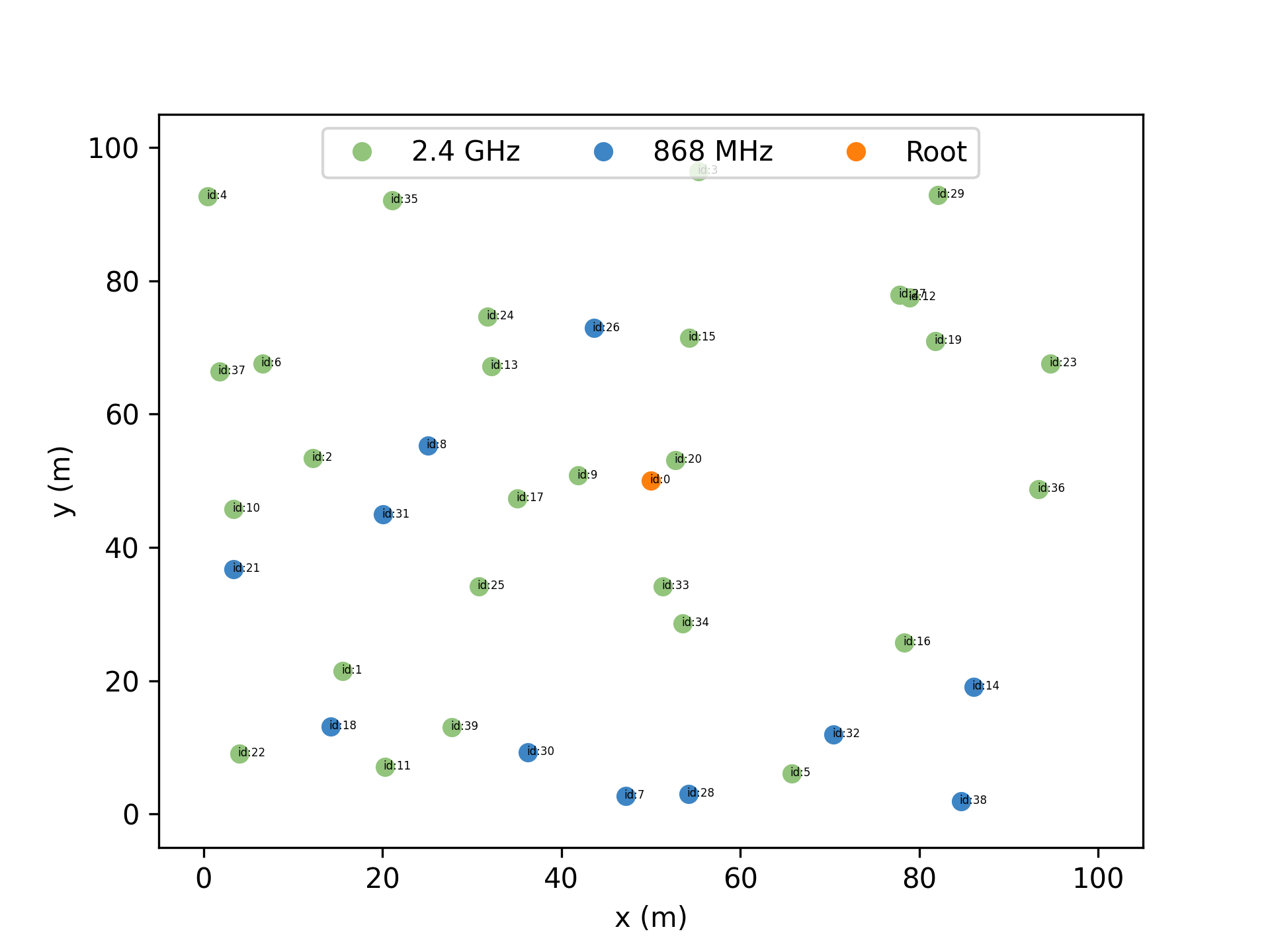}}
 \caption{Random deployment of 40 devices in 10km\textsuperscript{2} area highlighting the operating band that delivered the lowest average latency on each device. \label{fig_40_ran_deploy_avg_latency}}
\end{figure}

{We select a randomly deployed scenario with 40 nodes to better illustrate these findings. Table~\ref{tab_40random_retries} shows the total network packet retries and the average packet retransmissions by node, Figure~\ref{fig_cdf_lat_40_ran} presents the latency Cumulative Distribution Function (CDF), both for each operating band and the resulting combined network}. 
{It can be seen from Table~\ref{tab_40random_retries} that the 868 MHz band required approximately only 47\% of the number of retransmissions carried out by the 2.4 Ghz interface. Moreover, it is very interesting to note that the case of combined 868 MHz and 2.4 GHz interfaces required only 72\% of the retransmissions carried out by the 2.4 GHz interface when utilized alone, what has positive implications in  power consumption, for instance.}{ Furthermore, the similarity between combined and 2.4GHz Latency CDF curves from 0 to 1s in Figure~\ref{fig_cdf_lat_40_ran} indicates that the faster transmission times and shorter slotframe duration are the major players providing the lowest latency paths towards the DAG root. {However, as illustrated in Figure~\ref{fig_40_ran_deploy_avg_latency}, it is interesting to note that 11 out of the 39 nodes had a lower average latency in the 868 Mhz band, representing 28\% of the nodes. This shows that the sub-GHz band also helps in reducing the average network latency.}}



\begin{table}
\caption{Packet retries for the scenario with 40 nodes randomly deployed.}
\centering
\setlength{\tabcolsep}{3pt}
\begin{tabular}{|l|c|c|c|}
\hline
Metric & 2.4 Ghz & 868 Mhz & Combined \\
\hline
Network total & 506 & 237 & 364\\
Average by node & 12.97 & 6.07 & 9.333\\
\hline
\end{tabular}
\label{tab_40random_retries}
\end{table}

The two operating bands offer distinct trade-offs regarding reliability and latency, while the combination of both can improve overall network performance for 6TiSCH networks. The following is a summary of the advantages and drawbacks of the proposed scheme:

\begin{itemize}
\item Advantages: a) increases PDR by adding diversity and utilizing sub-Ghz bands that offer greater robustness against propagation losses; b) reduces latency by selecting the first received packet from either operating band; {c) reduces the amount of retransmissions, decreasing congestion, latency and consumption}. 
\item Drawbacks: a) requires a second radio interface at the nodes; b) increases the energy consumption as all packets are transmitted in a redundant form; c) increases application complexity in the border router or application server to manage and discard duplicated packets. 
\end{itemize}

\section{Conclusion and Future Directions} \label{sec_conclusions}

This work proposed the addition of a second radio interface in network devices to operate a redundant 6TiSCH network in sub-GHz bands. Results show an increase of  25\% in PDR and closely to 30\% decrease in latency in some scenarios. Observed gains are associated with diversity from a redundant transmission and physical link characteristics present in the sub-GHz band, which lead to an increased reliability and longer range. 



In future works the proposed method could be tested in large-scale topologies with hardware heterogeneity and in real environments. 
Moreover, dynamic radio selection could be investigated in combination with IPv6 packet tagging allowing diverse application QoS levels. Finally, additional studies are required to better comprehend the effects of TSCH and RPL networks regarding packet queue lengths and transmissions retries to prevent network bottlenecks. 

\bibliographystyle{IEEEtran}
\bibliography{IEEEabrv,artigo}

\end{document}